\documentclass[12pt,twoside,fleqn]{article}
\usepackage{epsfig}
\usepackage{multirow}
\usepackage{exscale}
\def\lsim{\raise0.3ex\hbox{$<$\kern-0.75em\raise-1.1ex\hbox{$\sim$}}}
\def\gsim{\raise0.3ex\hbox{$>$\kern-0.75em\raise-1.1ex\hbox{$\sim$}}}

\newcommand{\beqn}{\begin{equation}}
\newcommand{\eqn}{\end{equation}}
\newcommand{\bqa}{\begin{eqnarray}}
\newcommand{\eqa}{\end{eqnarray}}
\newcommand{\bqas}{\begin{eqnarray*}}
\newcommand{\eqas}{\end{eqnarray*}}
\newcommand{\bdm}{\begin{displaymath}}
\newcommand{\edm}{\end{displaymath}}

\begin{document}
\thispagestyle{empty}
%
 \mbox{} \hfill BI-TP 2001/15\\
\begin{center}
{{\large \bf The Chiral Critical Point in 3-Flavour QCD}
 } \\
\vspace*{1.0cm}
{\large F. Karsch, E. Laermann and Ch.  Schmidt}

\vspace*{1.0cm}
{\normalsize
$\mbox{}$ {Fakult\"at f\"ur Physik, Universit\"at Bielefeld,
D-33615 Bielefeld, Germany}
}
\end{center}
\vspace*{1.0cm}
\centerline{\large ABSTRACT}

\baselineskip 20pt

\noindent
We determine the second order endpoint of the line of
first order phase transitions, which occur in the light quark mass
regime of 3-flavour QCD at finite temperature, and analyze
universal properties of this chiral critical point.
A detailed analysis of Binder cumulants and the joint
probability distributions of energy like and ordering-field like
observables confirms that the chiral critical point belongs to the
universality class of the three dimensional Ising model.
From a calculation with improved gauge and staggered fermion actions
we estimate that the transition is first order for pseudo-scalar meson
masses less than about 200 MeV.
\vfill
\noindent
\mbox{}July 2001\\
\eject
\baselineskip 15pt

\section{Introduction}

In the chiral limit the order of the QCD phase transition
depends on the number of quark flavours that become massless.
It has been conjectured \cite{PisWi} and verified in numerical
calculations \cite{Gav87,Kog88,Bro90} that this phase transition is
first order for QCD with $n_f \; \ge \; 3$ massless quark flavours.
The strength of this first order transition
weakens in the presence of non-vanishing quark masses and
eventually ends in a second order endpoint. The analysis of effective
models constructed in terms of the chiral order parameter suggests
that this endpoint, the chiral critical point, belongs to the
universality class of the three dimensional Ising model \cite{Gav94}.
In fact, the entire surface of second order phase transitions which
in the $n_f$-dimensional space of quark masses separates the regime
of first order transitions at small quark masses from the
crossover region is expected to belong to this universality class.

In the case of degenerate quark masses the lattice formulation of QCD
depends on two bare couplings, the bare quark mass $m$ and
the gauge coupling $\beta \equiv 6/g^2$. The bare quark mass acts like
an external symmetry breaking field and thus leads to an explicit
breaking of the $SU_L(n_f)\times SU_R(n_f)$ chiral symmetry.
For $n_f \; \ge \; 3$ there is a first order phase transition in
the chiral limit ($m\equiv 0$), which will continue to persist for
small but non-zero values of $m$ up to a critical value, $\bar{m}$, of the
quark mass.
It is expected that the universal properties of this chiral critical
point are controlled by a global $Z(2)$ symmetry, which however is not
an obvious global symmetry of the QCD Lagrangian. It
rather is the relevant symmetry of the effective Hamiltonian which
controls the critical behaviour at this point.
Although the transition is second order at the chiral critical point, it
is obvious that neither the chiral condensate nor the Polyakov loop will
be an order parameter for the spontaneous $Z(2)$ symmetry breaking
at this critical point. In fact, we do not
know a priori what are the relevant observables representing
the energy-like and ordering-field like operators of the effective
Hamiltonian and the corresponding scaling fields (couplings) that define
the energy-like and ordering-field like directions in the coupling parameter
space; it is part of the problem of analyzing the universal
behaviour at this endpoint that one has to identify these operators
and couplings appropriately.

The problem of determining the critical properties at the endpoint
of a line of first order phase transitions is well known from studies
of other statistical and field theoretic models. The construction of
appropriate scaling fields has been discussed in detail in the case
of the liquid-gas phase transition \cite{wilding}. The concepts
developed in this context have recently also been used to locate
and explore the properties of the critical endpoint of the
electro-weak phase transition \cite{higgs1,higgs2} as well as
the critical point in the ferromagnetic, three dimensional 3-state Potts
model \cite{stickan}. The latter problem, of course, is closely related
to the analysis of the endpoint of the line of first order phase
transitions that occur in the heavy quark mass limit of QCD \cite{deF}.

We will focus here on a discussion of the light quark mass regime of QCD. In
particular we concentrate on an analysis of universal properties of the chiral
critical point in 3-flavour QCD. Although we will eventually also be interested
in an accurate determination of the location of this endpoint in terms
of physical mass values, e.g. quark or hadron masses, our main
emphasis here will be on the analysis of universal properties. We
thus will mainly use standard (unimproved) Wilson gauge and staggered fermion
actions for our analysis. Lattice calculations will be performed
on lattices with temporal extent $N_\tau=4$ and various spatial lattice
sizes to use well known finite size scaling methods at the critical point.
A first estimate for the location of the critical endpoint on lattices
with temporal extent $N_\tau=4$ has been given in \cite{Aoki99}. We confirm
this estimate here and will, in addition, present a detailed analysis
of the universal properties of the transition at the chiral
critical point. Using improved gauge and staggered fermion actions
we then will give a first estimate of the pseudo-scalar
meson mass at the chiral critical point.

This paper is organized as follows. In the next section we introduce
the basic variables relevant for the description of critical
behaviour in the vicinity of the chiral critical point. In section 3
we locate the chiral critical point and determine its universality
class using Binder cumulants. In section 4 we construct the
order parameter and energy-like operator at the chiral critical point.
Section 5 is devoted to the determination of the critical pseudo-scalar
meson mass at the critical point. Finally we give our conclusions
in Section 6.

\section{Scaling fields at the chiral critical point}

The thermodynamics of QCD with $n_f$ quark flavours is described
in terms of the partition function

\begin{equation}
Z(\beta, m) = \int {\cal D}U e^{-S(\{ U\}, \beta, m)} \quad ,
\label{partition}
\end{equation}
which is defined on a 4-dimensional lattice of size $N_\sigma^3\times
N_\tau$. Here $\beta = 6/g^2$ and $m$ denote the gauge coupling and
bare quark mass, respectively. The Euclidean action, $S$, is given
in terms of the pure gauge action $S_G$ and the fermion matrix
$Q_F$. Throughout this paper we will use staggered fermion actions
and standard Wilson or tree level improved gauge actions. The
discretized QCD action then reads,

\begin{equation}
S(\{ U\}, \beta, m) = \beta S_G(\{ U\}) - {n_f \over 4 } {\rm Tr}\;
\ln Q_F(\{ U\}, m) \; .
\label{action}
\end{equation}

\begin{figure}
\begin{center}
\epsfig{file= 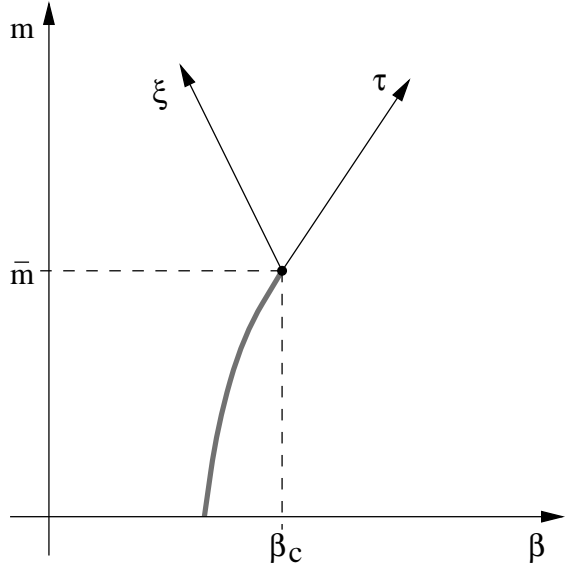,width=75mm}
\end{center}
\caption{Generic phase diagram in the low mass regime of 3-flavour QCD.}
\label{fig:generic3}
\end{figure}
The generic phase diagram of QCD in the small quark mass region
is shown in Fig.~\ref{fig:generic3} in the plane of bare couplings
appearing in the QCD partition function.
In the vicinity of the second order endpoint,
$(\beta_c(\bar{m}),\bar{m})$,
the dynamics and the universal critical behaviour is controlled by an
effective Hamiltonian, which can be expressed in terms of
two operators $\cal E,~M$, {\it i.e.} the energy-like and
ordering-field like operators that couple to two relevant scaling
fields $\tau$ and $\xi$,

\begin{equation}
{\cal H}_{eff} (\tau, \xi)  = \tau {\cal E} + \xi {\cal M} \quad .
\label{heff}
\end{equation}
Under renormalization group transformations the multiplicative
rescaling of the couplings $\xi$ and $\tau$ is controlled by the
two relevant eigenvalues that characterize the universal
critical behaviour in the vicinity of the second order critical
point. The singular part of the free energy density thus scales like

\begin{equation}
f_s(\tau, \xi) = b^{-3} f_s(b^{y_t}\tau, b^{y_h}\xi)
\quad ,
\label{fss}
\end{equation}
where the dimensionless scale factor $b\equiv LT =N_\sigma / N_\tau$
gives the spatial extent of the lattices in units
of the inverse temperature.
Susceptibilities constructed from $\cal E$ and $\cal M$ will show
the standard finite size scaling behaviour

\begin{equation}
\chi_{\cal E}\equiv V^{-1} \langle (\delta {\cal E})^2 \rangle
\sim a_{\cal E} b^{\alpha /\nu}
\quad , \quad
\chi_{\cal M}\equiv V^{-1} \langle (\delta {\cal M})^2 \rangle
\sim a_{\cal M} b^{\gamma /\nu}
\quad .
\label{fssEM}
\end{equation}
Here $V=N_\sigma^3$ denotes the $3$-dimensional spatial volume and
$\delta X \equiv X- \langle X \rangle$.
Note that $\cal E$ and $\cal M$ as well as $S_G$ and $\bar{\psi}\psi$
are defined to be extensive quantities.

For finite values of the quark mass all global symmetries of the QCD
Lagrangian are explicitly broken. As the symmetry of ${\cal H}_{eff}$
that characterizes the critical behaviour at the chiral critical point
is not shared in any obvious way by the QCD Lagrangian we also may expect
that in the vicinity of the chiral critical point the operators appearing
in the QCD Lagrangian are mixtures of the energy-like (${\cal E}$) and
ordering-field like (${\cal M}$) operators. Similarly the couplings
appearing in the QCD Lagrangian are linear combinations of the scaling
fields as we have indicated in Fig.~\ref{fig:generic3}.

In the vicinity of the critical point one may use a linear ansatz for
the couplings
\begin{eqnarray}
\tau &=& \beta - \beta_c +A\; (m - \bar{m}) \quad ,\nonumber \\
\xi &=& m - \bar{m} + B\; (\beta - \beta_c) \quad ,
\label{tauxi}
\end{eqnarray}
as well as for $\cal E$ and $\cal M$ which are constructed in terms
of operators appearing in the original QCD Lagrangian,
\begin{eqnarray}
{\cal E} &=& S_G + r\; \bar{\psi} \psi \quad , \nonumber \\
{\cal M} &=& \bar{\psi} \psi + s \; S_G \quad .
\label{EM}
\end{eqnarray}
Here $\bar{\psi} \psi$
denotes the chiral condensate evaluated on a given gauge field
configuration $\{ U\}$. In terms of the staggered fermion matrix,
$Q_F(m)$, this is given by
$\bar{\psi} \psi \equiv 0.25 \; n_f\; {\rm Tr}\; Q_F^{-1}(\{ U\}, m)$.

As the operators of the QCD Lagrangian,
e.g. $S_G$ and $\bar{\psi} \psi$ or related observables like the Polyakov
loop expectation value, are mixtures of $\cal E$
and $\cal M$, the corresponding susceptibilities will all receive
contributions from fluctuations of $\cal E$ as well as $\cal M$.
Asymptotically therefore all of them will
show identical finite size scaling behaviour which will be dominated
by the larger of the two exponents $\alpha /\nu$ and
$\gamma /\nu$, respectively. For the symmetry groups of interest in the
QCD context, e.g. the symmetry of three dimensional $Z(2)$ or $O(N)$
spin models, this will be $\gamma /\nu$. A finite size scaling
analysis of susceptibilities constructed from the basic operators
of the QCD Lagrangian thus will give access only to the ratio $\gamma /\nu$,
which unfortunately is quite similar for all the above mentioned
symmetry groups and thus is not a good indicator for the universality
class controlling the critical behaviour in the vicinity of the chiral
critical point.

The situation is different for cumulants constructed from linear
combinations of $\bar{\psi}\psi$ and $S_G$,
\begin{equation}
B_4(x) \; = \; {\langle \bigl( \delta  M (x)\bigr) ^4\rangle \over
\langle \bigl( \delta  M (x)\bigr) ^2\rangle^2 } \quad , \quad
M(x) = \bar{\psi}\psi + x\; S_G \quad .
\label{binder}
\end{equation}
From Eq.~\ref{fss} it follows that for arbitrary values of $x$ the
cumulants are renormalization group invariants which in the infinite
volume limit take on a universal
value at the critical point $(\tau, \xi)\equiv (0,0)$. For all values of
$x$ different from $1/r$ the cumulants behave asymptotically like the
Binder cumulant for the order parameter; cumulants calculated on
different size lattices for different quark masses will intersect at some
value of the quark mass. In the infinite volume limit these intersection
points will converge to a universal value which
is characteristic for the universality class of the
underlying effective Hamiltonian and, in fact, is quite different for
the classes of three dimensional $Z(2)$ and $O(N)$ symmetric spin models;
e.g. $B_4 = 1.604$ for $Z(2)$ \cite{b4ising}, 1.242(2) for $O(2)$ \cite{b4o2}
and 1.092(3) for $O(4)$ \cite{b4o4}.
The cumulants $B_4(x)$ thus seem to be appropriate observables to locate
the chiral critical point as well as to determine its universality
class without knowing in detail the correct scaling fields.

\section{Locating the chiral critical point}

The determination of the chiral critical point proceeds in
two steps. First of all, we determine for fixed values of the quark mass
pseudo-critical couplings, $\beta_{pc}(m)$, on finite lattices. These are
defined as the position of maxima in susceptibilities of $\bar{\psi}\psi$,
the gauge action $S_G$ and the Polyakov loop $L$ \cite{Kar94}.
We then make use of the finite size scaling properties of Binder cumulants
$B_4(x)$ evaluated at $\beta_{pc}(m)$. When analyzed as function
of the bare quark masses the cumulants calculated on
lattices with spatial extent $L_1$ and $L_2$ will intersect at a
mass $m_{L_1,L_2}$. For $(L_1,L_2)\rightarrow (\infty,\infty)$ these
intersection points will converge to the chiral critical point.

From previous studies with standard Wilson gauge and staggered fermion
actions one knows that the endpoint in 3-flavour QCD
is located close to $m=0.035$ \cite{Aoki99}.
As the universal properties of the endpoint are not expected to be
influenced by lattice cut-off effects we took advantage of this
knowledge and performed our detailed scaling analysis with unimproved
actions. We have performed calculations on lattices of size
$N_\sigma^3\times 4$, with $N_\sigma = 8$, 12 and 16. We have used
four values of  the quark mass in the interval $m \in [0.03, 0.04]$
and for each of these masses we calculated thermodynamic observables for
3 to 4 different values of the gauge coupling $\beta$.
In general we collected for each pair of couplings $(1-3)\cdot 10^4$
configurations generated with the hybrid-R
algorithm\footnote{We used trajectories of length $\tau = 0.675$ generated
with a discrete step size $\delta \tau = 0.015$.}.
Interpolations between results from different $\beta$-values have been
performed using the Ferrenberg-Swendsen multi-histogram technique \cite{Fer88}.
\begin{table}[htb]
\begin{center}
\begin{tabular}{|c|c|ccc|}
\hline
$m$ & $N_\sigma$
& $V^{-1}\left< ( \delta\bar\psi \psi)^2 \right>$
& $V^{-1}\left< ( \delta L           )^2 \right>$
& $V^{-1}\left< ( \delta S_G         )^2 \right>$ \\ \hline
0.0300&  8 & 13.4(2)   & 1.32(3)  & 3.21(10)      \\
      & 12 & 32.5(1.4) & 3.08(12) & 7.34(31)      \\
      & 16 & 65.0(4.7) & 6.27(44) & 14.8(1.0)     \\ \hline
0.0325&  8 & 12.9(4)   & 1.32(4)  & 3.21(12)      \\
      & 16 & 50.5(3.0) & 5.00(29) & 11.72(70)     \\ \hline
0.0350&  8 & 11.9(5)   & 1.34(5)  & 3.05(14)      \\
      & 12 & 23.5(1.5) & 2.52(16) & 5.73(40)      \\
      & 16 & 39.8(2.8) & 4.19(28) & 9.52(64)      \\ \hline
0.0400& 12 & 16.8(8)   & 2.05(11) & 4.45(24)      \\
      & 16 & 23.4(2.1) & 2.84(23) & 6.03(54)      \\
\hline
\end{tabular}
\end{center}
\vspace*{-0.4cm}
\caption{Volume dependence of the susceptibilities}
\label{tab:peaks}
\end{table}

In Tab.~\ref{tab:peaks} we summarize our results for the peak
heights of the three different susceptibilities which we have analyzed.
We generally find that the positions of these peaks coincide within
statistical errors. Thus in Tab.~\ref{tab:couplings} only the
pseudo-critical couplings extracted from the location of the peak
in $\langle (\delta \bar{\psi}\psi)^2\rangle$ are given. Also given in this
table is the Binder cumulant $B_4(x)$ calculated at the pseudo-critical
couplings for two values of $x$.
We have checked that the cumulants indeed attain their
minimum at $\beta_{pc}(m)$. The case $x=0$ corresponds to the
cumulant of the chiral condensate alone, whereas $x=0.43$ corresponds
to our best estimate for the mixing parameter $s$, whose determination
we are going to discuss in the next section.

\begin{table}[htb]
\begin{center}
\begin{tabular}{|c|c|c|cc|}
\hline
$m$ & $N_\sigma$ & $\beta_{pc}$ & $B_4(0)$ & $B_4(0.43)$ \\ \hline
0.0300&  8 & 5.1403(5)  & 1.637(37) &  1.610(37)        \\
      & 12 & 5.1411(4)  & 1.524(53) &  1.518(61)        \\
      & 16 & 5.1396(1)  & 1.454(87) &  1.453(87)        \\ \hline
0.0325&  8 & 5.1456(6)  & 1.623(51) &  1.607(49)        \\
      & 16 & 5.1458(2)  & 1.535(69) &  1.537(82)        \\ \hline
0.0350&  8 & 5.1524(5)  & 1.640(45) &  1.625(44)        \\
      & 12 & 5.1508(5)  & 1.664(55) &  1.657(53)        \\
      & 16 & 5.1499(1)  & 1.72(10)  &  1.72(10)         \\ \hline
0.0400& 12 & 5.1598(4)  & 1.896(63) &  1.889(63)        \\
      & 16 & 5.1593(5)  & 2.02(12)  &  2.01(12)         \\
\hline
\end{tabular}
\end{center}
\caption{Critical couplings and fourth order cumulants}
\label{tab:couplings}
\end{table}

\begin{figure}
\begin{center}
\epsfig{file= 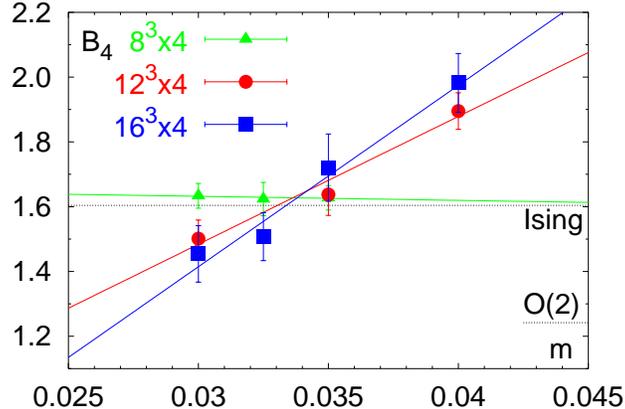,width=85mm}
\end{center}
\caption{The Binder cumulant of the chiral condensate,
$B_4 (0) = \langle (\delta \bar{\psi} \psi)^4\rangle /
\langle (\delta \bar{\psi} \psi)^2\rangle^2$.}
\label{fig:binder}
\end{figure}

The cumulant of the chiral condensate, $B_4(0)$, is shown in
Fig.~\ref{fig:binder}.
We note that the cumulants calculated on different size lattices
intersect at a quark mass close to $m=0.035$. The  value of $B_4(0)$
at the intersection point is compatible
with the universal value of the Binder cumulant for the 3-dimensional
Ising model. It is obvious from the data given in
Tab.~\ref{tab:couplings} that the situation is very similar for
$x=0.43$. We have fitted the cumulants on a given lattice size using a linear
ansatz in the quark mass, $B_4(x) = a_0 + a_1\; m$. From this we find
for the intersection point and the cumulant,
\vspace*{-0.4cm}\begin{eqnarray}
\hspace*{-0.5cm}x=0.0~:&
(\beta_c,\; \bar{m})\; =\; (5.1458(5),\; 0.0331(12)) \; , \;
B_4 (0) = 1.639 \pm 0.024 \; , \cr
\hspace*{-0.5cm}x=0.43~:&
(\beta_c,\; \bar{m})\; =\; (5.1454(5),\; 0.0329(15)) \; , \;
B_4 (x) = 1.624 \pm 0.023 \; .
\label{mcest}
\end{eqnarray}
Here we have determined the critical coupling $\beta_c(\bar{m})$ from
a linear interpolation of the pseudo-critical couplings
on the $16^3\times 4$ lattice which are given in Tab.~\ref{tab:couplings}.

In addition to the Z(2) value for the Binder cumulant we show in
Fig.~\ref{fig:binder} also the result for 3-dimensional $O(2)$
symmetric spin models. As
the Binder cumulant depends quite sensitively on
the underlying symmetry at the critical point the
result found by us for the crossing point strongly
suggests that the chiral critical point indeed belongs to the
universality class of the 3-dimensional Ising model.
The ratio of critical exponents $\gamma/\nu$ on the other hand
is not sensitive to the universality class. As expected one finds,
however, that all three susceptibilities show identical finite
size scaling behaviour; for the quark mass closest to the
critical point, $m=0.035$, the ratio of susceptibilities
calculated on lattices of size $N_\sigma = 12$ and 16
takes on the value 1.69(23), 1.66(22) and 1.66(24) for
chiral, Polyakov-loop and action susceptibilities,
respectively. This corresponds to a ratio of critical exponents
$\gamma/\nu =1.8(5)$ which is consistent with the 3-dimensional
Ising as well as $O(2)$ and $O(4)$ values ($\gamma/\nu \simeq 1.96$).

\begin{figure}
\begin{center}
\epsfig{file= 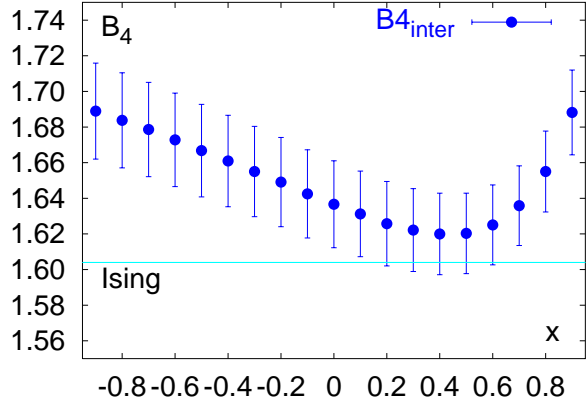,width=85mm}
\end{center}
\caption{The intersection point of Binder cumulants $B_4(x)$ calculated
on lattices of size $N_\sigma^3\times 4$ with $N_\sigma = 8$ and 16.
The horizontal line shows the universal Ising value.
}
\label{fig:intersect}
\end{figure}

As discussed above the intersection point of Binder cumulants
constructed from $M(x)$ according to Eq.~\ref{binder}
will be independent of the choice of $x$ in the infinite volume limit.
In fact, from Eq.~\ref{fss} it follows that for $x = s$ the singular part
of the free energy will lead to a unique crossing point, independent of the
lattice size, {\it i.e.} finite volume effects are minimized if we manage to
select the correct ordering-field like operator for constructing the Binder
cumulant.
For all other choices a volume dependence of the intersection points
calculated from Binder cumulants on different size lattices will result
from the scaling behaviour of $f_s(\tau, \xi)$ and they will converge to
the universal value only in the infinite volume limit.
It follows from Eq.~\ref{fss} that the volume dependence is quadratic
in $\Delta = x-s$, {\it i.e.} in deviations from the optimal choice
of $x$. This behaviour is confirmed through the analysis of
the $x$-dependence of the intersection point of the Binder cumulant
calculated on lattices of two different sizes.
In Fig.~\ref{fig:intersect} we show the intersection points of
Binder cumulants calculated on lattices of size $8^3\times 4$ and
$16^3\times 4$. As can be seen the volume dependence is small
for a wide range of $x$ values and the extremum, which corresponds
to the optimal choice $x=s$, is closest to the 3-d Ising value.
In fact, determining the $x$-value where the extremum is reached
provides a finite volume estimate for the mixing parameter $s$.
From a jackknife analysis one finds for the location of the minimum
$s_{\rm min} = 0.430(23)$.

\section{The mixing parameters $r$ and $s$}

Having located the chiral critical point we now can determine the
mixing parameters $r$ and $s$. This will allow us to construct
the energy- and ordering-field like operators and obtain further evidence
that the chiral critical point belongs to the universality class of
the 3-dimensional Ising model.

The parameters $r$ and $s$ are fixed by demanding that $\cal M$
should obey basic properties of the order parameter for
spontaneous symmetry breaking at the critical point.
In the symmetric phase the order parameter should stay constant
along the line of vanishing external field ($\xi \equiv 0$). Of course,
this should only hold in the vicinity of the critical point. We thus
demand
\begin{equation}
\biggl( { \partial \langle \cal{M} \rangle \over \partial \tau}
\biggr)_{|\tau = 0^+,\xi=0}
= 0\quad ,
\label{order}
\end{equation}
or equivalently
\begin{equation}
\langle \delta {\cal E} \delta {\cal M} \rangle  = 0 \quad .
\label{indep}
\end{equation}
Using these relations together with
Eqs.~\ref{tauxi} and \ref{EM} we obtain two
conditions for the mixing parameters $r$ and $s$ in terms of
the parameter $B$ and expectation values of $S_G$ and $\bar{\psi}\psi$,
\begin{eqnarray}
r &=& -\; B \quad , \nonumber \\
s &=& {\langle \delta \bar{\psi}\psi \delta S_G \rangle -B
\langle (\delta \bar{\psi}\psi )^2 \rangle \over
\langle (\delta S_G )^2 \rangle - B \langle \delta
\bar{\psi}\psi\;  \delta S_G \rangle } \quad .
\label{rs}
\end{eqnarray}

The parameter $B$ controls the mixing of the gauge coupling and bare
quark masses needed to define lines of constant $\xi$. In the symmetry
broken phase one line of constant $\xi$ is known to us;
the line of first order phase transitions defines
the zero external field line ($\xi = 0$) of the effective Hamiltonian.
We thus can extract $B$ from the quark mass dependence of the
pseudo-critical couplings. Using Eq.~\ref{tauxi} one obtains
\begin{equation}
B^{-1} = - \biggr( {{\rm d}\beta_c(m) \over
{\rm d} m}\biggr)_{\bigl| \; m=\bar{m}} \quad .
\label{slopeB}
\end{equation}
Knowing $B$ we also know $r$ and can construct $s$ using Eq.~\ref{rs}.

In order to determine the mixing parameter $r$ from Eq.~\ref{slopeB}
we have to approximate the derivative by finite differences as we
can perform calculations of $\beta_c(m)$ only at a discrete
set of quark mass values. A first estimate may be given using our
data on the largest lattice ($16^3 \times 4$).
The slope of $\beta_{pc}(m)$
defines $r^{-1}$. We estimate this from a straight line fit to
the values given in Tab.~\ref{tab:couplings} which gives,
$r = 0.51 (2)$. The mixing parameter $r$ is large and definitely non-zero;
as expected, a mixture of two operators is needed to construct an
energy like observable in which the otherwise dominant ordering-field
like contributions to $S_G$ and $\bar{\psi}\psi$ cancel. The importance
of choosing the correct
mixing parameter $r$ becomes apparent from an analysis of joint
probability distributions for $\delta {\cal E}$ and ${\delta \cal M}$.
These are little affected by changes of $s$ but they strongly
depend on $r$. In Fig.~\ref{fig:jointEM} we show contour plots for
the joint probability distributions and various values of $r$ and $s$.
For $r=s=0$ these are just the fluctuations in $\delta S_G$
and $\delta \bar{\psi}\psi$ which are strongly correlated. Only for
larger values of $r$ fluctuations in the order parameter are independent
of changes in the energy-like observable. In fact, we have used this
criterion to improve our determination of $r$ over our previous estimate
based on the slope of $\beta_c(\bar{m})$ at $\bar{m}$.
We demand that fluctuations in $\delta {\cal M}$ vanish for any
fixed value of $\delta {\cal E}$. This generalizes Eq.~\ref{indep}
and maximizes the $Z(2)$ symmetry of the contour plots shown in
Fig.\ref{fig:jointEM} around the $\delta {\cal M}=0$ axis. As an
estimate for the mixing parameter $r$ one obtains in this way
$r\; =\; 0.550(7)$. A determination of the parameter $s$
from Eq.~\ref{rs} then yields $s \; =\; 0.41(51)$. Although this value
has large errors it is consistent with the result found from the
extremum of intersection point of the Binder cumulant shown in
Fig.~\ref{fig:intersect}. As a best estimate of the mixing parameters
we therefore obtain
\begin{equation}
r\; =\; 0.550\pm 0.007 \quad, \quad s \; =\; 0.430\pm 0.023 \quad.
\label{bestmix}
\end{equation}

\setlength{\unitlength}{1mm}
\begin{figure}
\begin{picture}(120,140)
\put(43,138){\parbox{2cm}{$r=0.00$}}
\put(93.75,138){\parbox{2cm}{$r=0.55$}}
\put(5,105.5){\parbox{2cm}{$s=0.00$}}
\put(5,54){\parbox{2cm}{$s=0.43$}}
\put(50.75,51.5){\epsfig{file=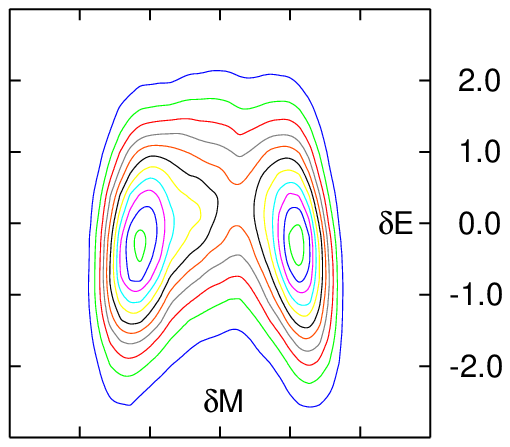,width=100mm}}
\put(50.75,0){\epsfig{file=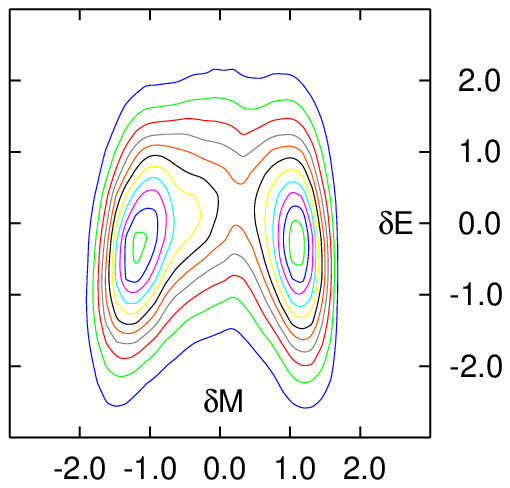,width=100mm}}
\put(0,51.5){\epsfig{file=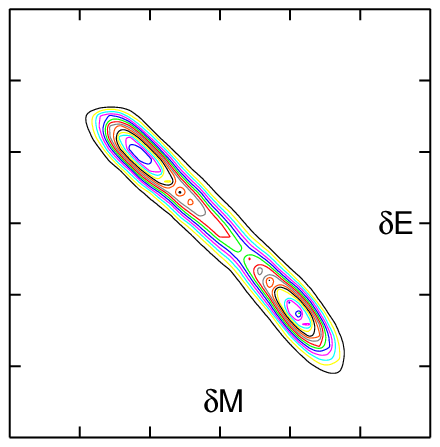,width=100mm}}
\put(0,0){\epsfig{file=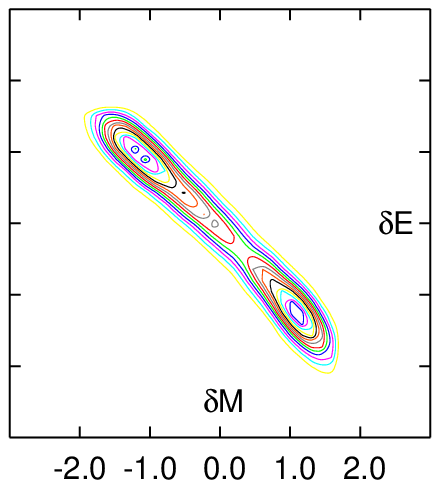,width=100mm}}
\end{picture}

\vspace*{0.6cm}
\caption{Joint probability distributions of the fluctuations in the
ordering-field like and energy like operators, which have been contructed
according to Eq.~\ref{EM}.
}
\label{fig:jointEM}
\end{figure}

\section{Physical scale at the chiral critical point}

The analysis presented in the previous sections confirms that for
finite values of the three light quark masses the chiral critical line
in QCD belongs to the universality class of the 3-dimensional Ising
model\footnote{This line may end in a tricritical point for some
finite value of the strange quark mass \cite{Gav94}.}. For the
case of three degenerate quark masses we have located the critical
point as given in Eq.~\ref{mcest}. In order to determine a
physical scale for this endpoint we have calculated
hadron masses on a $16^4$ lattice at $(\beta_c, \bar{m})$. For
the pseudo-scalar and vector meson mass we find $m_{\rm ps}=0.463(1)$
and $m_{\rm v}= 1.387(38)$, respectively. Expressing the pseudo-scalar mass
in units of the critical temperature, $m_{\rm ps}/T_c = 1.853(1)$,
and using estimates for the critical temperature in 2 and 3-flavour
QCD \cite{Kar00} we thus estimate
for the pseudo-scalar meson mass at the chiral critical point
$m_{\rm ps} \simeq 290$~MeV.

The entire analysis of the chiral critical point discussed so far
has been performed with unimproved gauge and fermion actions on
rather coarse lattices. Improved actions are not expected to
modify the results on the universal properties of the
chiral critical point, which have been presented above. They may, however,
well influence the quantitative determination of the chiral
critical point. It has been found in studies of the first order
deconfinement transition occurring in the pure gauge sector that
the gap in physical observables like the latent heat or surface tension
is cut-off dependent. Improved actions generally lead to smaller
gaps and a reduced cut-off dependence of these observables
\cite{Bei97}. One thus may expect that also in the region of first order
chiral transitions
the gap in the chiral condensate gets reduced when calculated
with improved actions. This will shift the critical point to
smaller values of the pseudo-scalar meson mass.
We therefore have investigated the chiral critical point also
in calculations with improved gauge and staggered fermion actions
(p4-action), which we have used previously for studies of the
thermodynamics of two and three flavour QCD
\cite{Kar00,Pei99}. In these calculations no evidence for a first
order transition has been found down to bare quark masses $m=0.01$
\cite{Pei99}. We now have extended these calculations to smaller
quark masses and made use of the universal properties of the Binder
cumulants discussed above. One can locate the chiral critical point
quite accurately through a calculation
of Binder cumulants on finite lattices and there is no need to
accurately determine the correct order parameter for such an analysis;
{\it i.e.} the analysis can be performed with Binder cumulants constructed
from the chiral condensate. We have performed calculations on a lattice
of size $12^3\times 4$ with a bare quark mass $m=0.005$
and on a lattice of size $16^3\times 4$ with $m=0.01$. Again a
Ferrenberg-Swendsen reweighting is used to determine the pseudo-critical
couplings and Binder cumulants at these couplings. The results are
summarized in Tab.~\ref{tab:improved}.

\begin{table}[t]
\begin{center}
\begin{tabular}{|c|c|c|c|c|}
\hline
$m$ & $N_\sigma$ & $\beta_{pc}$ & $B_4(0)$ &
$V^{-1}\langle (\delta \bar{\psi}\psi)^2\rangle$ \\ \hline
0.005&  12 & 3.2443(8)  & 1.31(12) & 54.8(5.6)   \\ \hline
0.01 &  16 & 3.2778(8)  & 2.14(10) & 11.5(1.0)   \\
\hline
\end{tabular}
\end{center}
\caption{Critical couplings, Binder cumulant of the chiral condensate
calculated and chiral susceptibility calculated on $N_\sigma^3\times 4$
lattices with the p4-action \cite{Kar00,Hel99}.}
\label{tab:improved}
\end{table}

From the results given in Tab.~\ref{tab:improved} it is obvious that
the smaller quark mass leads to a Binder cumulant below the value for
the 3-dimensional Ising model and thus is in the first order region
of the 3-flavour phase diagram. This also is confirmed by the large
value of the chiral susceptibility. For the larger quark mass, on the
other hand, the susceptibility stays small and the Binder cumulant is
significantly above the Ising value. This quark mass still lies in the
crossover region. We thus have obtained an upper and
lower limit for the chiral critical point, suggesting a
critical bare quark mass of $\bar{m}=0.0075(25)$.
Extrapolating the meson masses calculated in \cite{Kar00} to this
value of the bare quark mass, we estimate for pseudo-scalar meson
mass at the chiral critical point $m_{\rm ps} \simeq 192(25)$~MeV.
The critical mass thus is considerably smaller than estimated from
calculations with unimproved gauge and fermion actions.

\section{Conclusions}

Through an analysis of Binder cumulants we have verified that
the chiral critical point in three flavour QCD belongs to the
universality class of the three dimensional Ising model.
The analysis of joint probability distributions provides a
powerful tool to construct the order parameter at the chiral
critical point as well as the energy like scaling field. Although
the chiral condensate itself is not the order parameter at this
critical point, we explicitly have verified that Binder cumulants
constructed from it are little influenced by finite volume effects
and are good observables to locate the critical point as well
as the universality class.

Having determined the universality class of the chiral critical
point for three degenerate quark masses one could use this information
to determine the critical parameters
also for non-degenerate quarks from calculations of
Binder cumulants on finite lattices. The quality of the straight
line fits shown in Fig.~\ref{fig:binder} suggests that a first
order Taylor expansion of $B_4(0)$
in terms of degenerate up/down quark masses $m_{u,d}$ and
a strange quark mass $m_s$ around the three flavour critical point
$\bar{m}=0.033$ might be possible. As the chiral critical line corresponds
to those sets of quark masses where the Binder cumulant attains the
3-d Ising value one has to determine the line on which
$B_4(0)$ stays constant in the $(m_{u,d}, m_s)$-plane.
To leading order this line is given by

\begin{equation}
m_s = \bar{m} - 2\; (m_{u,d} - \bar{m})
\quad .
\label{criticalline}
\end{equation}
For instance, for $m_{u,d} = 0.025$ this estimate suggests that the
region of first order chiral transitions ends at a critical value
of the strange quark mass $m_s \simeq 0.049$. This is consistent with
the result of \cite{Bro90} where no sign for  a first order transition
has been found for $(m_{u,d},m_s)= (0.025,0.1)$ and not in
contradiction with \cite{Aoki99} who reported on a ``weak first order
like behaviour'' at $(m_{u,d},m_s)= (0.025,0.05)$ as opposed to two
state signals at lower $m_s$ values.

The chiral critical point has been determined in calculations with
unimproved as well as improved gauge and staggered fermion actions on
lattices with temporal extent $N_\tau=4$. The physical scale extracted
from calculations of the pseudo-scalar meson mass at these endpoints
is quite different in both cases. This indicates that cut-off effects are
still significant and calculations closer to the continuum limit
are definitely needed to fix a physical scale for the location
of the chiral critical point. The present indication is, however, that
improved actions lead to smaller values for the critical pseudo-scalar
meson mass.
This makes it increasingly unlikely that the transition in the
physically realized case of two light and a heavier strange quark
lies in the first order region of the QCD phase diagram.

\noindent
{\bf Acknowledgements:}

\medskip
\noindent
The work has been supported by the TMR network
ERBFMRX-CT-970122 and by the DFG under grant Ka 1198/4-1.


\end{document}